\documentstyle[12pt]{article}
\begin{document}
\newcommand{\newc}{\newcommand}

\newc{\be}{\begin{equation}}
\newc{\ee}{\end{equation}}
\newc{\ba}{\begin{eqnarray}}
\newc{\ea}{\end{eqnarray}}
\newc{\bann}{\begin{eqnarray*}}
\newc{\eann}{\end{eqnarray*}}
\newc{\ie}{{\it i.e. }}
\newc{\eg}{{\it e.g. }}
\newc{\etc}{{\it etc. }}
\newc{\etal}{{\it et al. }}

\newc{\ra}{\rightarrow}
\newc{\lra}{\leftrightarrow}
\newc{\no}{Nielsen-Olesen }
\newc{\lsim}{\buildrel{<}\over{\sim}}
\newc{\gsim}{\buildrel{>}\over{\sim}}
\begin{titlepage}
\begin{center}
December 1994\hfill       MIT-CTP-2398, BROWN-HET-983,\\
\vspace{-0.1cm}
\hfill SISSA Ref. 2/95/A
\vskip 0.1in
{\large \bf
Formation of Vortices in First Order Phase
Transitions
}\vskip .1in
{\large Alejandra Melfo}\footnote {E-mail address:
melfo@galileo.sissa.it.}\\[.10in]
{\em International School for Advanced Studies,(SISSA-ISAS)\\
     Strada Costiera 11, 34014 Trieste, Italy,}\\
and\\
{\em Laboratorio de F\'{\i}sica Te\'orica\\ Facultad de Ciencias,
Universidad de Los Andes\\ M\'erida 5101-A, Venezuela}\\[.15in]
 \vskip .1in
{\large Leandros Perivolaropoulos}\footnote{E-mail address:
leandros@mitlns.mit.edu}
\vskip .1in
{\em Center for Theoretical Physics, Massachussets Institute of Technology\\
77 Massachussetts Avenue\\
Cambridge, Mass. 02139, USA.}\\
and\\
{\em Department of Physics, Brown University,\\ Providence, R.I. 02912,
USA}\\[.15in]
\end{center}
\begin{abstract}
\noindent
Using a toy model Lagrangian we investigate the formation of vortices in first
order phase transitions. The evolution and interactions of vacuum bubbles are
also studied using both analytical approximations and a numerical simulation of
scalar field dynamics. A long lived bubble wall bound state is discovered and
its existence is justified by using a simplified potential for the bubble wall
interaction. The conditions that need to be satisfied for vortex formation by
bubble collisions are also studied with particular emphasis placed on
geometrical considerations. These
conditions are then implemented in a Monte Carlo simulation for the study of
the probability of defect formation. It is shown that the probability of vortex
formation by collision of relativistically expanding bubbles gets
reduced by about $10$\% due to the above mentioned geometric effects.
 \end{abstract}

\end{titlepage}
\setcounter{footnote}{0}
\section{Introduction}
Symmetry breaking phase transitions in the early universe can give rise to
topologically stable localized energy concentrations known as {\it topological
defects} (for recent reviews see Refs. \cite{hk94,p94,b93} ). These defects
can be points (monopoles), lines (cosmic strings) or surfaces (domain walls)
depending on the homotopy of the vacuum manifold of the broken phase
\cite{r87}.

 In general when a symmetry group $G$ is spontaneously broken to a smaller
group $H$, defects can form if the resulting vacuum manifold $M=G/H$ has
non-trivial homotopy. Cosmic strings (vortices in two space dimensions) form
when the first homotopy group of $M$ is non-trivial \ie  $\pi_1 (M) \neq 1$.

Consider for example the Lagrangian density describing the dynamics of a
complex
scalar field $\Phi = \Phi_1 + i \Phi_2$
\be
{\cal L} =  {1\over 2} \partial_\mu \Phi^* \partial^\mu \Phi - V(|\Phi|)
\label{lagrangian}
\ee
where $V(|\Phi|)$ is minimized for $|\Phi| = \sigma \neq 0$
(\eg $V(\Phi)= {\lambda \over 4} (|\Phi|^2 - \sigma^2)^2$). The set
of potential minima of $V(|\Phi|)$ ($\Phi=\sigma e^{i \alpha}$) has the
 topology of a
circle $S^1$. In a cosmological setup, according to the Kibble mechanism
\cite{k76}, there
will be (by causality) field configurations that span the whole vacuum
manifold
as we go around a large circle in physical space (\ie asymptotically $\Phi
\rightarrow
\sigma e^{i \theta}$ where $\theta$ is the azimuthal angle in physical space).
 Such configurations will inevitably form \cite{yb89} (but see Ref. \cite{rs93}
  for potential
loopholes in the gauge case), with probability about
$1\over 4$ \cite{lp91}, when  causally disconnected domains of the universe
merge as the causal horizon expands. The asymptotic behavior $\Phi \rightarrow
\sigma e^{i \theta}$ implies by continuity of $\Phi$ that there will be a
point inside the large circle
where $\Phi = 0$. This point (and its neighborhood) being outside of the vacuum
manifold will be associated with topologically trapped energy density.
This configuration is
the topologically stable vortex \cite{no73}. Extended to three dimensions this
object becomes a line defect, the {\it  cosmic string}.

In the above simplified Lagrangian (\ref{lagrangian}) no gauge fields are
 involved and the
broken symmetry is a $U(1)$ global symmetry leading to the formation of {\it
global vortices}. The price to pay for considering a simple $U(1)$ global
rather than a $U(1)$ gauge symmetry \cite{no73} is that the total energy of
an isolated
global vortex diverges logarithmically. This however is not a problem in
systems where a physical cutoff scale is built in, like multivortex systems
where the cutoff scale is the intervortex separation, or cosmological setups
where the cutoff is the horizon scale. For simplicity in what follows we will
consider sytems with global rather than gauge symmetry breaking.

The above picture of string formation when regions of the size of the horizon
at
the phase transition become causally connected, is characteristic of systems
undergoing a second order phase transition (\cite{vv84}, see \cite{bd94}
 for a recent discussion).
 In systems undergoing first order phase transitions, vortices can form by the
merging of expanding vacuum bubbles with scalar field phases such that the
whole vacuum manifold is covered after the bubble collision and phase
interpolation\footnote{Subcritical bubbles \cite{gg94} can also play a minor
role in vortex formation \cite{s92} but those effects are ignored within our
approximation}.
Consider for example an area $\Delta A$ of a two
dimensional system undergoing a first order phase transition during a time
interval $\Delta T$. There are three basic conditions that need to be satisfied
for a vortex to form within the area $\Delta A$ during the time $\Delta T$.
\begin{enumerate}
\item
The nucleation of at least three bubbles must take place during $\Delta T$.
\item
The nucleated bubbles must have phases such that geodesic interpolation leads
to complete coverage of the vacuum manifold $S^1$.
\item
The initial geometric configuration of the three nucleated bubbles must be such
that the collision of all three bubbles occurs before the phase interpolation
process can spoil the previous condition.
\end{enumerate}
Thus, the probability for a vortex to form within $\Delta A$ during the time
interval $\Delta T$ may be written as
\be
P_{tot} (\Delta A, \Delta T) =  B^3
P_{phase}(B) S(B,v)
\ee
where $B\equiv \Gamma {{\Delta A} \over {A_{bub}}}\Delta T $ is the probability
for a true vacuum bubble of area $A_{bub}$ to form within $\Delta A$ during
time $\Delta T$,
$P_{phase}(B)$ is the
probability that
a geodesic interpolation of the bubble phases completely covers the vacuum
manifold. The dependence on the probability $B$ exists because  $P_{phase}(B)$
is larger for clustered defects and therefore it increases with $B$.
The  suppression factor $S(B,v)$ depends
on bubble formation probability and the bubble wall expansion velocity $v$, and
is the probability that the initial configuration of bubbles will be such that
the third condition is satisfied.

The factor $P_{phase}(B)$ has been calculated in previous studies and found to
be between $0.25$ and $0.42$ depending on the number of clustered vortices.
 One of the main goals of this work is to find the suppression factor $S(B,v)$
for relativistically expanding bubbles ($v\simeq 1$). The case when bubbles
expand in the presence of plasma \cite{t92} ($v < 1$)
is significantly more complicated \cite{kibtalk} and will be included in
 a separate publication\cite{mpinpro}.

In the next section we derive and solve
numerically (and analytically in the thin bubble wall limit) the instanton
equations for bubble formation. We  use a simple scalar field potential with
vacuum manifold $S^1$ describing a first order phase transition. The obtained
scalar field configurations are then evolved by using a numerical simulation
based on a second order accurate leapfrog algorithm \cite{nr}. We first focus
on the bubble interactions and study their dependence on the phase difference
 between the two bubbles, both numerically and in the context of a simple
 analytic model.
 After justifying analytically the numerically observed existence of long lived
states of repelling walls in the limit of large phase difference, we show that
no metastable states exist in the model considered. Thus all colliding bubbles
eventually merge and the existence of metastable embedded walls need not be a
consideration for studying vortex formation. The formation of relativistic
phase
waves after bubble collisions is also seen in our simulations, confirming
expectations from previous studies.

In section 3,  we focus on the probability of
vortex formation during three bubble collisions. We first briefly discuss the
factor $P_{phase}(B)$ and show how can its dependence on the vacuum decay rate
$\Gamma$, which has been ignored in most previous studies, be used to explain
previous numerical results\cite{s92} that seemed puzzling at the time of their
derivation. We then focus on the geometric suppression factor $S(B,v\simeq 1)$
and first derive the geometrical conditions under which the vortex formation is
suppressed. These conditions are then tested by using dynamical simulations of
three bubble collisions, and a Monte Carlo simulation is constructed based on
these conditions to obtain the dependence of the suppression factor $S$ on the
mean interbubble distance and therefore on the bubble formation probability
$B$. Finally, in section 4 we conclude and briefly discuss extensions of this
work that are currently in progress

\section{Bubble evolution and interactions}

 Consider a complex scalar field $\Phi = |\Phi | e^{i\alpha}$, in a
(2+1)-dimensional spacetime,  whose dynamics is determined by the
Lagrangian (\ref{lagrangian}) and a symmetry-breaking potential
\be
V= \lambda \left[ {|\Phi |^2 \over 2} (|\Phi | - \sigma)^2 - {\epsilon \over 3}
\sigma |\Phi |^3 \right]
\label{potential}
\ee
with $\epsilon\ge 0$ (Fig. 1).

The false vacuum ($\Phi = 0$) of the potential (3) decays via bubble nucleation
to the true
vacuuum ($|\Phi | = \sigma \eta \equiv \sigma {1 \over 4}
( 3 + \epsilon + \sqrt{ (3+\epsilon)^2 -8 } )$). The two dimensional
field configurations of the bubbles
nucleated during this first order phase transition can be obtained \cite{c77}
(see also
\cite{b84} for a review) by
 solving the Euclidean field equations
\be
{\partial^2 \Phi \over \partial \rho^2} + {2\over \rho}{\partial\Phi \over
\partial\rho} = {\partial V(|\Phi|) \over \partial|\Phi|}
\label{euclid}
\ee
with $\rho^2 = |\vec x|^2 + \tau^2$, $\tau$ being euclidean time.
The initial configuration of the field after
tunneling has therefore an O(3) symmetry. This symmetry
 of the initial bubbles will become O(2+1) symmetry in Minkowski spacetime,
where $\rho^2 = |\vec x|^2 - t^2$. Solutions
 satisfy the boundary conditions
\be
\phi \ra 0 \;  as \; \; \rho \ra \infty \; \; and \; \; \frac{\partial
 \phi}{\partial \rho} \ra 0\; \;  as \; \;  \rho \ra 0
\label{bound}
\ee
Analytic solutions to (\ref{euclid}) can be found in the so-called thin wall
 approximation, i.e. for $\epsilon \ll 1 $, when the energy difference between
the minima is much smaller than the height of the potential barrier. For the
 potential (3), the thin-wall solution is of the form \cite{l83}
\be
|\Phi| = {\sigma \over 2} \left[ 1 - tanh\left(\frac{\sqrt{\lambda}
\sigma}{2}
 (\rho - R_o) \right)\right]
\label{thin}
\ee
where $R_o = 1/(\sqrt{\lambda}\sigma\epsilon)$ is the bubble's initial radius,
found by minimizing the total energy.

In the general, or thick-wall case, solutions to (\ref{euclid}) with the
boundary conditions (\ref{bound})
can be obtained numerically using a relaxation technique \cite{nr}. An
arbitrary
constant phase is then asigned in the interior of each bubble and the resulting
configurations are evolved by solving the dynamical field equations
\be
\ddot \phi - \nabla^2 \phi = - {\partial V \over \partial\phi}
\ee
where the dimensionless
 variables
\be
\phi = \Phi / \sigma \;, \hspace{0.5cm} \vec x = \vec{ \rm x} \,
m \;, \; \; \;
and \hspace{0.5cm} t = {\rm t} \, m
\label{dimension}
\ee
 with $ m = \sqrt{\lambda}\sigma $, were used.
For this part of the code, a second order leapfrog scheme \cite{nr} was
implemented on a 400x400  lattice, using as a reference the algorithm of
Ref. \cite{yb89}.   Energy was
conserved in all simulations to within $5$\% for the evolution timescales.

Defining the bubble walls as $\rho = R_o$, we see that the O(3+1) symmetry
forces
 the bubbles to expand, approaching the speed of light on a timescale
  determined
  by the initial radius $R_o$ ($|\vec x_{bub}|^2= R_o^2 + t^2$). Because of
   Lorentz
 contraction, thick-wall bubbles rapidly become thin-walled ones. As a
 consequence,
 interactions between bubbles occur almost always in the thin-wall regime,
but we have kept thick-wall bubbles as initial conditions for our simulations
 for
 generality.
For the figures presented, we have used $\epsilon = 0.8$ unless
specified otherwise.

It is evident from the simulations that the two-bubble interactions
 are strongly influenced by the difference between the field phases. Walls of
  bubbles with phase differences $\Delta \alpha \sim \pi$ tend to repell each
  other at short distances, thus delaying the time of merging of the walls.
  As bubbles approach, this  causes
the walls to separate and come in contact again, producing an oscillating
false-vacuum wall, as can be seen in Fig. 2. This domain wall eventually
decays,
 its lifetime depending on the phase difference. The effect is clearly observed
 when $\Delta \alpha \gsim 0.9 \pi$.
  For phase differences of $\Delta \alpha \simeq \pi$, we have found lifetimes
  of 10, (in the dimensionless units of Eq. (\ref{dimension})). Points in the
   wall
  located in the line joining the bubble's  center oscillate with a period
  $T\simeq 4$ ($\epsilon = 0.8$).
  Notice that this oscillatory
state occurs  {\it before} the bubbles merge, and
 is not to be confused with the oscillations of the field's magnitude that
 occur during bubble merging and after the phase interpolation has occured,
 as  reported in
 \cite{hms82} and
\cite{hdb94}. The origin of the effect in this case is the balance
between the repulsive force originating from the phase difference between
colliding
 bubbles
and the attractive force of the scalar field potential. On the other hand, in
 the
oscillations  discussed in  Refs. \cite{hms82} and
 \cite{hdb94} only the attractive potential force is relevant since the
phase difference on opposite sides of the oscillating false vacuum is zero.

 To better understand the numerically observed long lived oscillating states we
 consider the problem in the planar
 approximation and  write the equations of motion of bubble walls located at
 $x =\pm \xi(t)$ as
\be
\sigma \ddot{\xi} = - {\partial W \over \partial \xi}
\ee
where $\sigma$ is the mass per unit area of the wall, and $W$  the static
interaction
 energy between the walls, per unit area.
 Writing the field as $\phi = |\phi| e^{i\alpha}$, we have (primes denoting
derivatives with respect to the spatial coordinate)
\be
W = \int_{-\xi}^{\xi} dx \left[{(|\phi |')^2 \over 2} + {|\phi |^2 (\alpha')^2
\over 2} + V(\phi) \right]
\label{energy}
\ee
As a first approximation,
consider $V(\phi)$, $|\phi|$ and $|\phi|'$ as constants
 in the region between the bubble walls
and $\alpha' = \Delta \alpha / 2\xi$. The force between the walls will be
\be
F \simeq  -V({\bar \phi}) - (|{\bar \phi'}|)^2 + {|{\bar \phi}|^2
(\Delta \alpha)^2
\over 4\xi^2}
\ee
where ${\bar \phi}$ and ${\bar \phi'}$ are close but not excactly equal to
zero.
That is, the difference in phases produces a repulsive term dominating at short
 distances.
This simple picture can be improved by considering the total field
 configuration as
a sum of two bubble configurations, $\phi = |\phi | e^{i\alpha}$, with
$|\phi |$ approximated by the sum of two thin-wall bubbles
\be
|\phi | = {\eta \over 2} \left[1 + tanh\left({x-\xi \over 2}\right)\right] +
{\eta \over 2} \left[1 - tanh\left({x+\xi \over 2}\right)\right]
\ee
and $\alpha'
= \Delta \alpha / 2\xi$. The resulting potential energy $W$ is plotted in
Fig. 3
for different values of $\Delta \alpha$ and $\epsilon$, as a function of
$\xi$. Reducing $\epsilon$ has the
effect of reducing the false vacuum potential energy, allowing the
 repulsive term
to dominate at greater distances. The effect of increasing
$\Delta \alpha$ is that
 of displacing the minimum of the potential to higher values of $\xi$. The
 approximation  is expected to be valid only for $\xi\gsim 1$ \ie before the
 bubbles begin to merge.
Our simulations have shown that the gradient is actually locally time dependent
 and the initially binding potential  of Fig. 3 eventually gets dominated by
 the attractive terms as the phase gradient decreases locally.

The existence of these oscillatory states leads naturally to the question of
existence
of ``embbedded'' meta-stable domain walls for this potential.
 Such domain walls  would be static solutions to the field equations
\ba
|\phi|'' - (\alpha')^2 |\phi| &=& {\partial V \over \partial|\phi|}\nonumber \\
\alpha'' + {2\over |\phi|}|\phi|'\alpha' &=& 0
\label{static}
\ea
with boundary conditions $\phi \ra 0$ as $x\ra 0$;  $\phi \ra \sigma \eta$ for
 $x\ra \infty$.
A simple asymptotic analysis suffices to prove that there is no solution
to (\ref{static}) satisfying the boundary conditions. As $x \ra 0$, we
can always write $|\phi| = C_{\phi} x^a$, $\alpha = C_{\alpha} x^b$. However,
inserting this in (\ref{static}) we  obtain
\be
a = {1 \over 2} \hspace{0.5cm}  b=-1 \hspace{0.5cm} C_{\alpha}^2 =
{-{1 \over 4}}
\ee
So the asymptotic form of the equations is not satisfied for a real coefficient
$C_{\alpha}$,
and therefore the solution with the required boundary conditions does not
exist.
Thus the only possible domain wall occurs for
$\Delta \alpha = \pi$, where the situation is analogous to that of a real
field.
Then Eqs. (13) become
\be
|\phi|'' ={\partial V \over \partial|\phi|}
\ee
with the boundary conditions $\phi = \pm \sigma \eta$ at
$x \ra \pm \infty$,  and $\phi = 0$ at $x=0$. The solutions are the domain
walls
in this potential. For $\epsilon$ nonzero, we can see that its existence is
 guaranteed by
noting, as usual, that the problem is analogous to that of a particle
moving in the
potential $-V(\phi)$, with the field representing the particle's position
and the spatial coordinate interpreted as a time coordinate. The ``particle''
is energetically allowed to start from  $x = + \sigma \eta$ at $t = -\infty$ ,
and arrive at $x = - \sigma \eta$ at $t = +\infty$.
This is not the case when $\epsilon = 0$, and then only solitons that go from
$x = \pm \sigma \eta$ to $x = 0$ are possible. But as we have seen, even in the
case where this domain wall does exist, the asymptotic analysis above means
that
  it is unstable under small variations of $\Delta \alpha$ (a static solution
  does not exist for $\Delta
\alpha \neq \pi$). As discussed above this was also verified by our
simulations.

After the two bubbles merge, the situation is the one already studied
numerically in \cite{hms82} and analytically in
\cite{hdb94}. A phase wave is generated in the
contact point, and expands inside the bubbles with the speed of light,
interpolating between the original phases.  The field
magnitude oscillates in the region of contact, with amplitude that depends
in the phase difference, this time being inversely proportional:
a greater phase difference will
result in a more energetic phase wave, that carries away the wall's energy more
efficiently thus dissipating the vaccum oscillations rapidly. These
oscillations
  however do not affect the phase interpolation:  the phase waves are produced
  as soon as the interiors of the bubbles come  into contact, and escape at
  light speed.

\section{Vortex Formation}

We turn now to discuss the probability of vortex formation in a three-bubble
collision. As noted in the introduction, the probability of forming three
bubbles in an area $\Delta A$ within a period of time $\Delta T$ is
$B^3$. In order to form a
vortex, the phase of these three bubbles must be distributed so that the
interpolation following collision leads to complete coverage of the
vacuum manifold (condition(2)). The probability for this to happen can
be estimated as follows \cite{lp91}:
Consider a triangular lattice in a 2d physical space and asign a random phase
to each point on the lattice. The question is {\it What is the probability
 that a vortex lies within a given triangle?} The mean phase difference of
 two neighbouring lattice points of a triangle is clearly $2\pi / 4$. In order
 that a vortex forms in that triangle, the whole vacuum manifold must be
  covered by interpolating the phases of the three lattice points and therefore
  the phase of the third lattice point of the triangle must lie on
the opposite part of the phase circle. Thus, on the average, the phase of the
 third lattice point  should be in a range of $2 \pi /4$. This will happen
 with probability $1/4$.
Thus, the probability for forming an isolated vortex in physical space is
$1/4$.
In realistic cases however vortices do not form isolated but in vortex
 antivortex
clusters especially when the vacuum
decay rate is high. A very relevant question
therefore is: {\it What is the probability for forming a vortex in a vortex
 antivortex cluster?}.
 The probability $p_{+-}$ for forming an antivortex next to an already formed
 vortex is larger than the probability for forming an isolated vortex. The
reason  is that the mean phase difference between neighboring lattice points
in a vortex
surrounding triangle is not $2 \pi /4$ but $2 \pi / 3$ and therefore to form an
antivortex next to a vortex we only need that the fourth lattice point be in a
phase range $2 \pi / 3$. So $p_{+-} = 1/3$. Thus, the probability for
having $i$ antivortices around a vortex is
\be
P_i = \left(\begin{array}{c}3\\ i \end{array}\right) (P_{+-})^i
(1 - P_{+-})^{3-i}
\label{pv-a}
\ee
and the probability per defect in clusters of four or larger is
$p_{+-}^{3/4}=0.43$
which is significantly larger than the naively obtained result of $0.25$.

This effect of  clustering explains the result found in Ref. \cite{s92},
where a simulation of the vortex formation process is done by placing
random-phase bubbles in a 2-dimensional lattice, and allowing them to
evolve and collide.
The author finds a probability of vortex and antivortex formation of 0.42 (10
 vortices and antivortices formed after the nucleation of 23 closely packed
  bubbles),
instead of the expected 0.25, and atributes his result to unknown dynamical
effects.
However, if we notice that the chosen rate of nucleation produces a highly
packed  system, and calculate the probability using (\ref{pv-a}), we obtain
0.43, in agreement with the simulations.

The dynamical delay of merging due to the phase repulsive potential, discussed
 in the previous section, does not affect vortex formation. The phase
interpolation will occur eventually, and the crucial factor is not the time it
takes, but the spatial range over wich the bubbles overlap, wich determines
the spatial sectors where phases will interpolate. A phase repulsion (which in
any case occurs for $\Delta \alpha$ extremely close to $\pi$, an unlikely
situation if the phases are required to be distributed in the vacuum manifold
as in condition (1) of
 the Introduction), can only delay the interpolation event.
 In the formation of vortices, it is only important to know {\it if} (not {\it
when}) the
phases of each of the three bubble pairs will geodesically interpolate.

In addition, for the case of bubbles with relativistic
 velocities (i.e., when the effects of the plasma are negligible), we
 have both the bubble walls and the phases propagating at $v\simeq 1$.
This means that the phase waves can never reach the bubble walls leading to a
single new
bubble with interpolated phase, before the third bubble has time to reach
the two collided ones.

Thus, in the absence of plasma,
the important ingredient
needed to find the  probability of vortex formation
is the initial geometric distribution and nucleation times of the colliding
 bubbles.
In the case when the effects of plasma are ignored ($v\simeq 1$) simulations
show (Fig. 4) that we can picture the bubbles as circles, even when the merging
 has occurred, the phase wavefront continuing the circle formed by the walls.

For a vortex to form, then, it is easy to see that condition (3) of section 1
is
satisfied {\it iff the three  expanding circles have at least one common
intersection point}. Thus, for example, the extreme case of  three bubbles of
 the
 same size (i.e., nucleated at the same time), but with centers located along
a straight line, will  not form a vortex.  If we allow for different nucleation
 times, bubbles will interact  having different sizes,  and then even
  non-aligned configurations  will not lead to a vortex (Fig. 5). Notice that
in contrast to what has been stated in previous  studies \cite{hdb94} it is in
 general not possible to select a frame where all bubbles nucleate at the same
time. The condition of simultaneity of three events is that there is a
{\it space-like} planar surface that goes through these events. This
is not always possible (consider for example
 the special case of three events on
a straight line \ie effectively in one spatial dimension). Even if it was
possible  to select a frame of simultaneity for three events in $2+1$
dimensions it would not be appropriate to do so in a Monte Carlo simulation
as this artificial selection of a frame would introduce a bias in the
measured probabilities.
 We therefore have used different times for the nucleation of each bubble.

The equation of motion for the bubble walls is dictated by the initial
 configuration,
as stated in the previous section, so that the radius of
a bubble nucleated at time $t=\Delta t_i$ will be $r_i(t)^2 = R_o^2 +
(t-\Delta t_i)^2$, the subindex $i$ going from 1 to 3. Given the positions
$(x_i, y_i)$
and nucleation times of three bubbles,
 one has only to solve the system of three equations for the intersection
 point and time
\be
(x - x_i)^2 + (y - y_i)^2 = R_o^2 + (t-\Delta t_i)^2
\label{condition}
\ee
A vortex forms if the solution to the system is real, positive and finite.

As a test for this condition, we have checked using the numerical simulation
 described in section 2  the formation of vortices and predicted formation
 times for several configurations. The geometrical condition of circle
  intersection was found to be accurate
in all cases, confirming the hypothesis that dynamical effects can be
ignored in the case of relativistically expanding bubbles.

Having tested the geometric model with dynamical simulations, a suppression
factor for vortex formation can now be found using a Monte Carlo  simulation.
 Random position and nucleation times were assigned for bubbles
 inside an area $\Delta A \equiv (\lambda R_o /2)^2$ and configurations were
 subject  to the requirement  that bubbles do not form in an overlapping state.
   After rescaling
   the considered area such that $R_o = 1$,  we constructed 500 randomly chosen
three-bubble geometrical configurations. The system (\ref{condition}) was
solved
 in each case and the number of cases with no triple intersection was counted.
We defined  the ratio of the number of cases where triple intersection occured
over the total number  of cases as the {\it suppression factor} for vortex
formation. This is the factor   $S(B,v\simeq 1)$ of Eq. 2. A plot of $S$ vs
 $\lambda \sim B$ is shown in Fig. 6.

Clearly the geometric suppression factor is less important
for high nucleation rate (small $\lambda$) but it is not negligible for low
 nucleation rates.

\section{Conclusion}

We have studied the basic
conditions that are needed for vortex formation by the merging of vacuum
bubbles
nucleated during first order phase transitions. There are three such conditions
 which include the existence of three colliding vacuum bubbles, the complete
coverage of the vacuum manifold by geodesic interpolation of the bubble phases,
 and the existence  of a triple collision point for the merging of the three
 circles that describe the relativistic expansion of the bubbles.
 The probability that each condition  is satisfied was obtained and the
 result was compared with previous studies. In particular  the existence of
a triple collision point during the evolution of three colliding
  bubbles occurs with probability approximatelly $92$\% for low vacuum
  decay rate.
Such a suppression of defect formation rate is not expected to modify in any
major way cosmological models based on cosmic strings.

We have considered the case of relativistically expanding bubbles and have
therefore neglected the friction effects of plasma particles surrounding the
expanding vacuum bubbles. Such particles, being massless in the false vacuum
 but massive in the true vacuum (inside the bubble), are expected to scatter
  on the bubble walls and decelerate them to non-relativistic velocities.
In models where the effects of plasma are important enough to lead to slowly
 expanding bubbles, our analysis gives only  an upper bound to the geometric
 suppression factor $S(B,v)$ which is expected to
 rapidly drop as the bubble wall velocity $v$ decreases.
Indeed, for small bubble wall velocity $v$, the interpolating phase front
propagating
{\it always relativistically} inside the bubbles after the first collision,
 is more efficient in
equilibrating the phases to the interpolated value before the third bubble
reaches the bubbles that collided first. Thus the formation of the vortex
 can be
avoided much more efficiently.  The study of the dependence of $S$ on $v$
in the
presence of plasma requires the detailed numerical simulation of the effects of
the plasma, and the construction of a generalized geometrical model and
 Monte Carlo simulation based on these effects. That work is currently in
  progress\cite{mpinpro}.

\vspace{0.5cm}
{\large \bf Acknowledgements}

We would like to thank Tanmay Vachaspati and Alex Vilenkin for
interesting discussions. We also thank Andrew Sornborger for his help with
 the numerics.
Special thanks are due to Robert Brandenberger for insightful comments and
discussion and for helping to make this collaboration possible.
Finally one of us (AM) is is grateful to the Physics Department of
Brown University
and SISSA-ISAS for
financial support. The numerical part of this project was
completed at the Center for Scientific Computing of Brown University.
Financial support for this project was provided in part by DOE Grants
DE-FG0291ER40688, Task A and DE-FC02-94ER40818.


\begin{thebibliography}{999999}
\bibitem{hk94} `{\it Cosmic Strings}'
M.B. Hindmarsh, T.W.B. Kibble, SUSX-TP-94-74, Nov 1994. 139pp.
Submitted to Rept.Prog.Phys.
(Bulletin Board: hep-ph@xxx.lanl.gov - 9411342).
\bibitem{p94} {\it Cosmic String Theory: The Current Status},
L. Perivolaropoulos,
Contributed to Summer School in High Energy Physics and Cosmology,
Trieste, Italy, 13 Jun - 29 Jul 1994.
\bibitem{b93}
R. Brandenberger,
{\it Int.J.Mod.Phys} {\bf A9}, 2117 (1994)
(Bulletin Board: astro-ph@babbage.sissa.it - 9310041).
\bibitem{r87}see e.g. R. Rajaraman, `{\it Solitons and Instantons}', North
Holland Publishing (1987), p. 78, and references therein.
\bibitem{k76} T.W.B. Kibble,
{\it J.Phys.} {\bf A9}, 1387 (1976).
\bibitem{yb89}
 J. Ye and  R. Brandenberger
{\it Mod.Phys.Lett.} {\bf A5}, 157 (1990).\\
J. Ye and  R. Brandenberger
{\it Nucl. Phys} {\bf B346}, 149 (1990).
\bibitem{rs93}S. Rudaz and A. M. Srivastava, {\it Mod.Phys.Lett.} {\bf A8}
1443 (1993)
(Bulletin Board: hep-ph@xxx.lanl.gov - 9212279).
\bibitem{lp91} R. Leese and T. Prokopec, {\it Phys. Lett.} {\bf B260}, 27
 (1991).
\bibitem{no73} H. B. Nielsen and P. Olesen, Nucl. Phys. B{\bf{61}}, 45 (1973).
\bibitem{vv84} T. Vachaspati and  A. Vilenkin,
{\it Phys.Rev.} {\bf D30}, 2036 (1984).\\
T. Vachaspati, {\it Phys.Rev.} {\bf D44}, 3723 (1991).
\bibitem{bd94}
R. Brandenberger and A. Davis {\it Phys.Lett.} {\bf B332} 305, (1994).
\bibitem{gg94}G. Gelmini and M. Gleiser, {\it Nucl. Phys.}
{\bf B419},129 (1994)
(Bulletin Board: hep-ph@xxx.lanl.gov - 9211303).
\bibitem{s92} A. M. Srivastava, {\it Phys. Rev.} {\bf D46}, 1353 (1992).
\bibitem{t92} N. Turok,
{\it Phys. Rev. Lett.} {\bf 68} 1803 (1992).\\ Bao-Hua Liu, Larry McLerran
and Neil Turok, {\it Phys.Rev.} {\bf D46},
2668 (1992).\\ M. Dine, R.Leigh, P. Huet, A. Linde and D. Linde,
{\it Phys. Rev.}
 {\bf D 46} 550 (1992).
\bibitem{mpinpro} A. Melfo and L. Perivolaropoulos in preparation (1995).
\bibitem{kibtalk}T.W.B. Kibble, talk presented at the Cambridge Workshop on
 Formation
and Evolution of Topological Defects (1994).
\bibitem{nr} {\it Numerical recipes in FORTRAN : the art of scientific
 computing}
                W. H. Press {\it et al.}, 2nd ed.
 Published by Cambridge [England] ; New York, NY, USA : Cambridge University
                Press (1992).
\bibitem{c77} S. Coleman,  {\it Phys. Rev.} {\bf D15}, 2929 (1977).
\bibitem{b84} R. Brandenberger {\it Rev.Mod.Phys.} {\bf 57}, 1 (1985).
\bibitem{l83}
A.D. Linde, {\it Nucl.Phys.} {\bf B216}, 421 (1983), ERRATUM-ibid.B223:544,
(1983). \\ A. D. Linde, {\it Particle physics and inflationary cosmology};
Published by Chur [Switzerland] ; New York : Harwood Academic Publishers,
(1990)
\bibitem{hms82} S. W. Hawking, I.G. Moss and J.M. Stewart, {\it Phys. Rev.}
{\bf D26}, 2681 (1982).
\bibitem{hdb94} M. Hindmarsh, A.C. Davis and R. Brandenberger,
{\it Phys. Rev.} {\bf D49} (1994).



\end{thebibliography}
\end{document}